\documentclass[11pt,aps,prd,reprint,nofootinbib]{revtex4}
\usepackage{epsfig}
\usepackage{graphicx}
\usepackage{amsfonts}
\usepackage{latexsym}
\usepackage{amsmath,amssymb}
\usepackage{verbatim}
\usepackage{mathtools}
\usepackage{setspace}
\usepackage{slashed}
\usepackage[all]{xypic}

\usepackage{tikz}
\usepackage{psfrag}
 \usepackage{subfig}
\usepackage{comment}
\usepackage{array}
\usepackage{amssymb}
\usepackage{amsmath}
\usepackage{amsthm}
\usepackage{graphicx}
\usepackage{float}

\begin{document}
	
\title{Second order causal hydrodynamics in Eckart frame : using gradient expansion scheme}
\author{Sayantani Lahiri}
\affiliation{University of Bremen, Center of Applied Space Technology and Microgravity (ZARM), 28359 Bremen}
\email{sayantani.lahiri@zarm.uni-bremen.de}

\begin{abstract}
In the present work, we develop a causal theory of relativistic non-ideal fluids up to the second order in the Eckart frame using gradient expansion scheme. Keeping the spirit of Mueller-Israel-Stewart formalism, the general forms of bulk viscosity, shear viscosity tensor and the heat flow vector are presented. Since each of the flux quantities explicitly carry curvature terms, we show that our formalism finds application in astrophysics in particular in the strong gravity regime. We elucidate two such applications namely in viscous thick accretion disks also known as Polish doughnuts and in addressing non-rotating equilibrium configuration, like neutron stars. 

\vspace{120mm}
	
\end{abstract}

\maketitle

\pagebreak
\section{Introduction}
Past many years, the framework of relativistic hydrodynamics has found applications in various disciplines like relativistic heavy ion collisions \cite{Heinz,W,Romatschke-3}, cosmology \cite{Zimdahl, Roy, Baier-cosmo} and astrophysics \cite{Font}.  
The theory of hydrodynamics was first formulated by Euler for ideal fluids \cite{Euler}, much later it was extended for non-ideal fluids by Navier and Stokes involving dissipative quantities like viscosity, heat flow, etc. \cite{Navier, Stokes}. 
Initial attempts to study the relativistic theory of fluids were made by Eckart \cite{Eckart} and by Landau-Lifshitz \cite{Landau} who developed the relativistic description of Euler equation for ideal fluids and Navier-Stokes for non-ideal fluids.
However, it was found that under linear perturbations, the relativistic version of Navier-Stokes equation allows acausal signal propagation and suffer from instabilities \cite{Hiscock}.
In particular, it was noted that the acausal behaviour of the Navier-Stokes equation was related to its parabolic nature which may be attributed to presence of first order gradients
as in the Navier-Stokes theory where each dissipative quantities is constructed of first order gradients of fluid hydrodynamical variables namely the fluid four-velocity, temperature and chemical potential. 
The dissipative quantities response instantaneously to the first order changes in the Eckart and Landau-Lifshitz approaches. 
Taking into consideration of theory of general relativity, later,  Muller \cite{Isr-1} and Israel-Stewart (MIS) \cite{Isr-2}, came up with a causal theory of relativistic dissipative hydrodynamics that was established to be stable under linear perturbations \cite{Hiscock-2}.
In contrary to Eckart formalism, the MIS theory is constructed using second order gradients of fluid hydrodynamical variables and its causality preserving property is closely related to the hyperbolic nature of the conservation laws in presence of second order gradient terms.
The theory of relativistic hydrodynamics is regarded as an effective theory in which the microscopic scale which constrains the validity of hydrodynamics is the mean free path ${\it l_{m}}$ such that ${\it l_{m}}<< L$, where $L$ is the macroscopic length scale of the system under consideration.    
In the effective theory description, the effects of non-ideal hydrodynamics appear as small deviations  from local thermodynamic equilibrium  which can be realized by appearances of order of gradients in the energy momentum tensor of the system. 
Thus in absence of fluctuations, the ideal fluid description described by the local equilibrium state corresponds to the zeroth order hydrodynamics, Navier-Stokes theory is the first order theory and the MIS formalism is the second order theory in gradients that gives rise to causal and stable theory of viscous hydrodynamics.
Then the gradient expansion scheme corresponds to systematic expansion in terms of gradients of fluid hydrodynamical variables in which the correction due to dissipative terms to the ideal fluid description are realized in terms of gradients. However the associated transport coefficients are the input parameters of the theory which are determined using an underlying microscopic theory e.g. kinetic theory.\\
A in-depth study of relativistic viscous fluids has been investigated using systematic expansion of the gradients of the hydrodynamical variables up to second order by Baier et.al. \cite{Baier} and Romatschke \cite{Romatschke-1} in the context of conformal and non-conformal theories respectively under the assumption of zero conserved charges in the Landau-frame with a primary motivation to address a number of phenomenon of relativistic heavy-ion physics.  
With a motivation to apply the causal theory of relativistic viscous hydrodynamics in field of astrophysics, where the heat flux is not negligible in presence of conserved charges, in the present work, we  develop the causal theory of relativistic non-ideal fluids in the Eckart frame using gradient expansion scheme up to second order.
Keeping the spirit of MIS theory, in section 2, we present the general forms of bulk viscosity, shear viscosity tensor and heat flow vector up to second order in gradients. In section 3, we put forward two possible applications of the formalism. First, in understanding shape of an accretion disk in a given black hole spacetime under the influence of shear viscosity and curvature tensor and
second, comprehending effects of bulk viscosity on a spherically symmetric non-rotating star under static equilibrium configuration.
We conclude our work with further discussions.

\section{Second order hydrodynamics in the Eckart frame }
The relativistic hydrodynamics of perfect fluids is a well-established theory \cite{Landau} and is discussed widely in the existing literature.
In the present paper
we choose therefore to include the relevant aspects that are necessary for us to construct the relativistic hydrodynamical theory of non-ideal fluids.
For simplicity, let us assume there exists a single species of conserved charge in the system characterized by its number density $n$. The constitutive relations of the ideal fluid in terms of the fluid four-velocity $u^{\mu}(x)$ are given by,
\begin{eqnarray}
N_{ideal}^{\mu} = n u^{\mu}, \qquad
T_{(0)}^{\mu \nu} =  e \, u^{\mu} u^{\nu} + p \,\triangle ^{\mu \nu} \label{current} 
\end{eqnarray} 
where $e$, $p$ are respectively the equilibrium total energy density and the equilibrium pressure of the fluid.
The ideal fluid hydrodynamics corresponds to the thermodynamic equilibrium configuration of the fluid with four-velocity $u^{\mu}=(\gamma,\gamma \overrightarrow{v})$ where $\gamma$ is the Lorentz factor and $p,e,n$ are slowly varying fields.   
In the curved spacetime, the projection tensor is defined as $\triangle ^{\mu \nu}=g^{\mu \nu}+u^{\mu}u^{\nu}$ with the property $u_{\mu}\triangle^{\mu \nu}=0$ and the normalization condition $u^{\alpha}u_{\alpha}=-1$. 
If the covariant derivative is decomposed as $\nabla^{\mu}=-u^{\mu}D + \mathcal{D}^{\mu}_\bot $ where the longitudinal and transverse parts are respectively $D=u^{\alpha}\nabla_{\alpha}$ and  $\mathcal{D}^{\mu}_\bot= \triangle^{\mu \nu} \nabla_{\nu}$, then using (\ref{current}), 
the conservation of the particle current $\nabla_{\alpha}N_{ideal}^{\alpha}=0$ and the conservation of the energy-momentum tensor $\nabla_{\alpha}T_{(0)}^{\alpha \beta}=0$ lead to the continuity equation, energy conservation equation and the Euler equation as follows, 
\begin{eqnarray}
Dn +n \nabla.u&=&0   \label{4},\\
D e +(e+p)\nabla.u &=&0,  \label{5}\\
a^{\mu}=Du^{\mu}& =&-\displaystyle \frac{1}{(e+p)}\mathcal{D}^{\mu}_\bot p \label{6}
\end{eqnarray}
In case of an ideal fluid, the constitutive relations do not contain any  derivatives/gradients of hydrodynamical variables so ideal fluid hydrodynamics is also known as zeroth order hydrodynamics. 
In presence of dissipation, the idea of global thermodynamic equilibrium breaks down and the ideal fluid description is no longer valid. 
As a result, the hydrodynamic variables are not constant leading to presence of non-zero fluxes in the local rest frame of the fluid. 
Under this situation, the hydrodynamical variables namely the four velocity of the fluid $u^{\mu}(x)$, temperature $T(x)$ and the chemical potential $\mu(x)$ are not uniquely defined in out-of-the equilibrium situations and therefore one can have several local values of the hydrodynamical variables at every point in spacetime, each differing by gradients but leads to their respective equilibrium values in absence of dissipation. 
This results into ambiguities in the definition of these variables in presence of dissipation. 
To get rid of ambiguities one resorts to a choice of frames.
There are in general two choices namely the Eckart frame (no charge diffusion) and the Landau frame (no heat flow). 
In the present paper, we will adopt the Eckart frame where we have $u_{\mu}N^{\mu}=n$, where $n$ is the conserved charge in the system. In presence of conserved charges one can refer to \cite{Muronga} for more discussions on relativistic viscous hydrodynamics.
The constituent relation for non-ideal fluids involving the most general form of energy momentum tensor and the particle current is given by,
\begin{eqnarray}
N^{\mu}=nu^{\mu}, \qquad
T^{\mu \nu}&=&e u^{\mu} u^{\nu}+ (p+\Pi)\triangle^{\mu \nu}+q^{\mu}u^{\nu}+q^{\nu}u^{\mu}+ \pi^{\mu \nu} \\
\vspace{2mm} &=& T_{(0)}^{\mu \nu}+T_{(non-ideal)}^{\mu \nu} \\
\text{where,}\qquad \, T_{(non-ideal)}^{\mu \nu}&=&\Pi\triangle^{\mu \nu}+q^{\mu}u^{\nu}+q^{\nu}u^{\mu}+ \pi^{\mu \nu} \label{non-ideal}
 \end{eqnarray}  
Here $q^{\mu}$ denotes the heat flow vector. $\Pi$ is the trace part and $\pi^{\mu \nu}$ is the symmetric transverse traceless part of the symmetric viscosity tensor. 
It is to be noted that in the Eckart frame characterized by no charge flow, the continuity equation remains unchanged as that of an ideal fluid and in presence of a single conserved charge the four-velocity is expressed as, $u^{\mu}= \displaystyle \frac{N^{\mu}}{\sqrt{-N^{\alpha}N_{\alpha}}}$.
The following two other conservation laws are respectively the Navier-Stokes equation and energy conservation,
\begin{eqnarray}
\mathcal{D}^{\mu}_\bot (p+\Pi)+(e+p+\Pi)a^{\mu}+\pi^{\mu \lambda}a_{\lambda}+\triangle^{\mu}_{\,\beta}\, \mathcal{D}_{\sigma}^\bot \pi^{\beta \sigma}+
\triangle^{\mu}_{\,\lambda}D q^{\lambda} +\left(\sigma^{\mu}_{\lambda}+\Omega^{\mu}_{\lambda} 
+\displaystyle \frac{4}{3}\triangle^{\mu}_{\lambda}(\nabla.u)\right)q^{\lambda}&=&0 \label{Navier-1} \nonumber \\
De+ (e+p+\Pi)(\nabla.u)-2\eta \sigma^{\mu \nu}\sigma_{\mu \nu}+q_{\alpha}a^{\alpha}+ \nabla_{\alpha}q^{\alpha}&=&0 \nonumber
\end{eqnarray}  
where the bulk viscosity, heat flow vector, shear viscosity and the vorticity tensor are respectively, 
 \begin{eqnarray}
 \Pi&=&-\zeta (\nabla.u),  \quad
 q^{\beta}=-\kappa T (\mathcal{D}^{\beta}_\bot \ln T + D u^{\beta}),\nonumber \\[1mm]
  \pi^{\mu \nu}& =&  -2 \eta \sigma^{\mu \nu} = - 2\eta \left(\displaystyle \frac{\mathcal{D}^{\mu}_\bot u^{\nu}+\mathcal{D}^{\nu}_\bot u^{\mu}}{2}-\displaystyle \frac{1}{3} \triangle^{\mu \nu}\nabla.u\right), \qquad  
  \Omega_{\alpha \beta}=\displaystyle \frac{1}{2} \left(\mathcal{D}_{\alpha}^\bot u_{\beta}-\mathcal{D}_{\beta}^\bot u_{\alpha}\right)  \label{visco-1}
  \end{eqnarray}
Here $\zeta\geq 0$, $\kappa \geq 0$ and $\eta \geq 0$ are the bulk viscosity coefficient, thermal conductivity and the shear viscosity coefficients. 
When each of the thermodynamic fluxes is expressed in terms of first order gradients of hydrodynamical variables so that the energy momentum tensor contains terms only up to
first order in gradients, such a theory is called the first order hydrodynamics.
But the relativistic Navier-Stokes equation and the heat conduction equation violate causality due to their parabolic nature allowing superluminal velocity of propagating signals \cite{Romatschke-1}, \cite{Roy} and the associated equilibrium states are plagued with instabilities \cite{Hiscock}. 
MIS theory which took into account of by general relativity, extended the method of Grad's 14-moment approximation and successfully rectified the causality violating nature of the conservation laws of relativistic non-ideal fluids by considering second order gradients of hydrodynamic variables. 
As a result the general forms of $\Pi$, $q^{\beta}$ and $\pi^{\mu \nu}$ contain additional coefficients arising due to inclusion of second order gradients. These additional coefficients are known as second order transport coefficients which serve as input parameters of the theory and are essentially evaluated using an underlying microscopic theory describing the system. 
Over the years, both first and second order transport coefficients have been widely computed in the context of both conformal, non-conformal theories and kinetic theory \cite{Romatschke-2,Muronga-1, Erdmenger,Moore,Saki,Kovtun}. 
Although the MIS formulation resulted into hyperbolic conservation laws devoid of causality and instabilities, it is not complete in the sense that this formalism missed out possible spacetime curvature terms at the second order, all of which are likely to exist if all possible combinations of second order gradients are considered.
In this direction, Baier, Romatschke et.al. \cite{Baier, Romatschke-2} reformulated IS theory by incorporating curvature terms in the Landau frame for both conformal and non-conformal theories with the
primary motivation to implement their approach in relativistic heavy-ion collisions.
However for astrophysical applications Eckart frame of reference is more favored, for example, a realistic accretion disk should have some radiation process which may be attributed to shear viscosity as well as heat flow. Therefore in presence of a single species of conserved charge, the effects heat flow is a non-negligible quantity and can only be addressed if the Eckart frame is considered.
Hence by adopting a similar approach taken in \cite{Baier, Romatschke-2}, we develop the general forms of bulk viscosity, shear viscosity and heat flow vector up to second order in gradients in order to obtain general relativistic causal hydrodynamical equations in the Eckart frame.\\
\textit{Procedure}: The conservation laws of ideal fluid eq.(\ref{5}) and eq.(\ref{6}) can be re-expressed using thermodynamical identities 
$e +p = Ts + \mu n$, $d e = T ds+ \mu dn$ and Gibbs-Duhem relation $dp=sdT+nd\mu$ as follows,
\begin{eqnarray}
D \ln s& =& -\nabla.u,  \label{A}\\
Du^{\mu}& =&- \mathcal{D}^{\mu}_\bot \ln T-\frac{n}{(s+\alpha n)}\mathcal{D}^{\mu}_{\bot} \left(\frac{\mu}{T}\right)  \label{B}
\end{eqnarray}
where $s$, $T$, $\mu$ are respectively the entropy density, temperature, the chemical potential of the conserved charge of the fluid and $\alpha= \displaystyle\frac{\mu}{T}$.
In the non-relativistic limit, $D \simeq \partial_{t}$ and $\mathcal{D}^{\mu}_{\bot} \simeq \partial_{i}$ (up to higher order gradients), so eq.(\ref{A}) and eq.(\ref{B}) indicates time derivatives can be recast into space derivatives which further implies not all first order gradients are independent.
In our approach, we take second order gradients of hydrodynamical variables i,e. four velocity $u^{\mu}$, temperature $T$ and chemical potential $\mu$ to construct scalars (for bulk viscosity), vectors orthogonal to $u^{\mu}$ (for heat flow vector) and symmetric traceless tensors orthogonal to $u^{\mu}$ (for shear tensor) in the Eckart frame to construct a general form of energy momentum tensor. As mentioned earlier, in absence of charge diffusion, the particle current is not modified in the Eckart frame.
\\
In a system with a single conserved charge, the first order  linearly independent components giving rise to second order corrections to energy momentum tensor are : a scalar $\mathcal{D}^{\alpha}_\bot u_{\alpha}$, two vectors $\mathcal{D}^{\nu}_\bot \mu $, $\mathcal{D}^{\nu}_\bot \ln T	$ and a tensor $\mathcal{D}^{\bot}_{ \alpha} u_{\beta}$.
Then all dissipative quantities up to second order of gradients are constructed out of linearly independent  structures consisting of comoving spatial first order gradients namely $\mathcal{D}^{\alpha}_\bot u_{\alpha}, \mathcal{D}^{\nu}_\bot \mu $ and $\mathcal{D}^{\nu}_\bot \ln T$ that appear in equations of motion given by eq.(\ref{A}) and eq.(\ref{B}) of the ideal fluid.
There are two considerations which we have used to obtain the second order theory using gradient expansion. 
First, the rank-two tensor $\mathcal{D}_{\alpha}^\bot u_{\beta}$ can be expressed in terms of trace part plus symmetric and antisymmetric trace-less parts orthogonal to $u^{\mu}$ as shown below,
\begin{equation}
\mathcal{D}_{\alpha}^\bot u_{\beta}=\displaystyle \frac{\nabla.u}{3}\triangle_{\alpha \beta}+\displaystyle \frac{1}{2}\sigma_{\alpha \beta}+\Omega_{\alpha \beta} \label{decomp-1}
\end{equation}
Second, the curvature term arises as a consequence of non-commutative property of covariant derivatives acting on the four-velocity vector $u^{\mu}$, which can be illustrated as follows,
	\begin{equation}
R^{\lambda}_{\:\mu \alpha \beta}\,u_{\lambda} =\nabla_{\alpha}\nabla_{\beta}\,u_{\mu}-\nabla_{\beta}\nabla_{\alpha}u_{\mu}
\end{equation}
where the Riemann tensor is : $R^{i}_{ \: jnl}=\displaystyle \frac{\partial}{\partial x^{n}}\Gamma^{i}_{\:lj}- \displaystyle \frac{\partial}{\partial x^{l}}\Gamma^{i}_{\:nj} + \Gamma^{s}_{\:lj}\Gamma^{i}_{\:ns}-\Gamma^{s}_{\:nj}\Gamma^{i}_{\:ls}$.\\[2mm] 
Considering all possible second order terms, then general forms of $\Pi$, $q^{\mu}$ and $\pi^{\mu \nu}$ at the second order can be presumed to be as follows,
\begin{eqnarray}
\Pi&=& -\zeta (\nabla.u)+{\large \Sigma}_{i=1}^{i=10} \alpha_{i}{\cal M}_{i} \label{1} \\
q^{\mu}&=&-\frac{\kappa n T^2}{(e+p)} \mathcal{D}^{\mu}_\bot \left(\frac{\mu}{T}\right)+\Sigma_{i=1}^{i=11} \beta_{i}{\cal N}^{\mu}_{i}  \label{2}\\[1mm]
\pi^{\mu \nu}&=&-2\eta \sigma^{\mu \nu}+\Sigma_{i=1}^{i=10} \lambda_{i}{\cal O}^{\mu \nu}_{i}  \label{3}
\end{eqnarray}
where $\alpha_{i}$, $\beta_{i}$ and $\lambda_{i}$ are coefficients at the second order.
In the following, we describe the details for obtaining eq.(\ref{1})- eq.(\ref{3}) using second order gradients.
The possible second order terms are given by,					
\begin{eqnarray}
&(\nabla.u)^2, \quad (\mathcal{D}^{\mu}_\bot \ln T) (\mathcal{D}^{\nu}_\bot \ln T),\quad (\mathcal{D}^{\mu}_\bot u^{\alpha}) (\mathcal{D}^{\nu}_\bot \ln T),\quad(\mathcal{D}^{\mu}_\bot u^{\alpha}) (\mathcal{D}^{\nu}_\bot u^{\beta}),  \nonumber\\  &R^{\lambda}_{\,\mu \alpha \beta}, \quad
\mathcal{D}^{\mu}_\bot \mathcal{D}^{\nu}_\bot u^{\alpha}, \quad \mathcal{D}^{\mu}_\bot \mathcal{D}^{\nu}_\bot \ln T,\quad \mathcal{D}^{\alpha}_\bot \mu \mathcal{D}^{\beta}_\bot \mu,\quad \mathcal{D}^{\alpha}_\bot \mathcal{D}^{\beta}_\bot \mu
\end{eqnarray}
which are used to construct possible scalars, vectors and tensors at the second order for constructing the dissipative flux quantities. Here the traceless symmetric part of a tensor is  $A^{<\mu \nu>}=\displaystyle{\frac{1}{2}}\triangle^{\mu \alpha}\triangle^{\nu \beta} (A_{\alpha \beta}+A_{\beta \alpha})-\displaystyle{\frac{1}{3}}\triangle^{\mu \nu}\triangle^{\alpha \beta}A_{\alpha \beta}$. For convenience, we denote each scalar by ${\cal M}_{i}$, each vector by ${\cal N}_{i}$ and each tensor by ${\cal O}^{\mu \nu}_{i}$, where $i$ labels a particular term.  In the following, we tabulate the possible terms,
\begin{table}[h]
	\caption{Possible scalars } 
	\centering 
	\begin{tabular}{l |l|l|l|l|l|} 
		\hline\hline 
		& ${\cal M}_1=\mathcal{D}_{\alpha}^\bot\mathcal{D}^{\alpha}_\bot \ln T$ & ${\cal M}_2=(\mathcal{D}^{\alpha}_\bot \ln T)\mathcal{D}_{\alpha}^\bot \ln T$ & ${\cal M}_3=(\nabla.u)^2$& ${\cal M}_4=R$& ${\cal M}_5=u^{\alpha}u^{\beta}R_{\alpha \beta}$\\[4pt]  
		&${\cal M}_6=\sigma^{\alpha \beta}\sigma_{\alpha \beta}$& ${\cal M}_7=\Omega^{\alpha \beta} \Omega_{\alpha \beta}$& ${\cal M}_8=(\mathcal{D}^{\alpha}_\bot \mu) \mathcal{D}_{\alpha}^\bot \mu$ & ${\cal M}_9=\mathcal{D}_{\alpha}^\bot\mathcal{D}^{\alpha}_\bot \mu$& ${\cal M}_{10}=(\mathcal{D}_\bot^{\alpha}\mu)\mathcal{D}_{\alpha}^\bot \ln T$  \\ [1pt] \hline 
	\end{tabular}
\end{table}
\begin{table}[h!]
	\caption{Possible vectors } 
	\centering 
	\begin{tabular}{l |l|l|l|l|} 
		\hline\hline 
		&  ${\cal N}_1^{\nu}=(\mathcal{D}_{\alpha}^\bot \ln T)\displaystyle \frac{(\nabla.u)}{3} \triangle^{\alpha \nu }$ & ${\cal N}_2^{\nu}=(\mathcal{D}_{\alpha}^\bot \ln T)\sigma^{\alpha \nu}$ & ${\cal N}_3^{\nu}=(\mathcal{D}_{\alpha}^\bot \ln T)\Omega^{\alpha \nu}$& ${\cal N}_4^{\nu}=(\mathcal{D}_{\alpha}^\bot \mu)\displaystyle \frac{(\nabla.u)}{3} \triangle^{\alpha \nu }$\\[5pt]
		& ${\cal N}_5^{\nu}=(\mathcal{D}^{\nu}_\bot \mu) (\nabla.u)$& ${\cal N}_6^{\nu}=(\mathcal{D}_{\alpha}^\bot \mu)\sigma^{\alpha \nu}$& ${\cal N}_7^{\nu}=\triangle^{\alpha \nu} u^{\gamma} R_{\alpha \gamma}$&${\cal N}_8^{\nu}=\mathcal{D}_{\alpha}^\bot \sigma^{\alpha \nu}$\\[5pt]
		& ${\cal N}_9^{\nu}=\mathcal{D}_{\alpha}^\bot \Omega^{\alpha \nu}$ & ${\cal N}_{10}^{\nu}=(\mathcal{D}_{\alpha}^\bot \mu)\Omega^{\alpha \nu} $ &${\cal N}_{11}^{\nu}= (\mathcal{D}_{\alpha}^\bot \mu)R^{\alpha \nu}$&\hspace{6mm}-- \\ [1pt] \hline 
	\end{tabular}
\end{table}
\begin{table}[h]
	\caption{Possible tensors } 
	\centering 
	\begin{tabular}{l |l|l|l|l|} 
		\hline\hline 
		& ${\cal O}^{\mu \nu}_{1}=(\mathcal{D}^{<\mu}_\bot \ln T) (\mathcal{D}^{\nu>}_\bot \ln T)$&
		${\cal O}^{\mu \nu}_{2}= \mathcal{D}^{<\mu}_\bot \mathcal{D}^{\nu>}_\bot \ln T$& $  {\cal O}^{\mu \nu}_{3}=(\mathcal{D}^{<\mu}_\bot \mu) (\mathcal{D}^{\nu>}_\bot \mu)$& ${\cal O}^{\mu \nu}_{4}=\mathcal{D}^{<\mu}_\bot \mathcal{D}^{\nu>}_\bot \mu$\\[5pt]
		&$ {\cal O}^{\mu \nu}_{5}=\mathcal{D}^{<\mu}_\bot \ln T \mathcal{D}^{\nu>}_\bot \mu$& ${\cal O}^{\mu \nu}_{6}=\sigma^{\gamma<\mu}\Omega^{\nu>}\,_{\gamma}$&
		${\cal O}^{\mu \nu}_{7}=\sigma^{\gamma<\mu}\sigma^{\nu>}\,_{\gamma}$& ${\cal O}^{\mu \nu}_{8}= \Omega^{\gamma<\mu}\Omega^{\nu>}\,_{\gamma}$ \\[5pt]
		&${\cal O}^{\mu \nu}_{9}= R^{< \mu \nu>}$& ${\cal O}^{\mu \nu}_{10}=u_{\alpha} u_{\beta} R^{\alpha<\mu \nu> \beta}$&\hspace{6mm}--&\hspace{6mm}--
		\\ [1pt] \hline 
	\end{tabular}
\end{table}
\newpage
It can be seen that eq.(\ref{1})- eq.(\ref{3}) are of algebraic form which violate causality and lead to unstable equations. 
In order to produce a causal as well as a stable theory in the similar spirit of Muller-Israel-Stewart formalism, dynamical equations for dissipative flux quantities are required to be considered by introducing relaxation-time coefficients for $\Pi$, $q^{\mu}$ and $\pi^{\mu \nu}$. 
We now present the general form of the shear viscosity tensor up to second order in gradients
\begin{eqnarray}
\pi^{\mu \nu} &=& -2 \eta \sigma^{\mu \nu} -\tau_{\pi} ^{<}D (-2 \eta)\sigma^{\mu \nu \, >}-2 \eta\tau_2\left[ \frac{D \eta}{\eta}  - \frac{(\nabla.u)}{4}   
+ \frac{D\ln T}{4} \right] \sigma^{\mu \nu} \nonumber \\
&&\, + \xi_2 \,\mathcal{D}^{<\mu}_\bot \mathcal{D}^{\nu>}_\bot \ln T + \xi_3\, (\mathcal{D}^{<\mu}_\bot \mu) (\mathcal{D}^{\nu>}_\bot \mu) + \xi_4\,\mathcal{D}^{<\mu}_\bot \mathcal{D}^{\nu>}_\bot \mu +\xi_5 \,\mathcal{D}^{<\mu}_\bot \ln T \mathcal{D}^{\nu>}_\bot \mu  \nonumber \\
&&\, + \xi_6 \,\sigma^{\gamma<\mu}\Omega^{\nu>}\,_{\gamma} + 
\xi_7 \,\sigma^{\gamma<\mu}\sigma^{\nu>}\,_{\gamma} +
\xi_8 \, \Omega^{\gamma<\mu}\Omega^{\nu>}\,_{\gamma} +
\kappa_1 \,	R^{< \mu \nu>}+
\kappa_2 \,u_{\alpha} u_{\beta} R^{\alpha<\mu \nu> \beta}
\label{Final-shear}
\end{eqnarray}
which will reproduce to a causal theory as explained in section III. Let us first show that eq.(\ref{Final-shear}) constructed using gradient expansion scheme reduces to the relaxation-type theory which was first introduced in MIS formalism to combat causality nature of first-order theories. 
The motivation is to express eq.(\ref{Final-shear}) in a form that will give rise to dynamical equations of $\pi^{\mu \nu}$ by keeping the essence of MIS theory. 
For this, we employ the similar procedure as employed in \cite{Romatschke-2} and \cite{Rischke}.
To construct a dynamical equation for $\pi^{\mu \nu}$ and in order to correctly reproduce a relaxation-type equation, a new basis is chosen (now with coefficients $c_i$) and the the combined term $^{<}D \sigma^{\mu \nu \, >} + T \nabla_{\lambda}\left(\frac{u^{\lambda}}{4 T}\right) \sigma^{\mu \nu}$ \footnote{This particular combination consisting of second order terms is present in MIS theory for $\pi^{\mu \nu}$.} is expressed in the new basis with ${\cal{O}}_{i}^{\mu \nu}$ as follows,
	\begin{eqnarray}
	^{<}D \sigma^{\mu \nu \, >} + T \nabla_{\lambda}\left(\frac{u^{\lambda}}{4 T}\right) \sigma^{\mu \nu}
	=c_{1}{\cal{O}}_{1}^{\mu \nu}+c_{2}{\cal{O}}_{2}^{\mu \nu}+........ c_{9}{\cal{O}}_{9}^{\mu \nu}   \label{a}
	\end{eqnarray}
	where $c_{i}$'s are coefficients of the new basis.
 We now eliminate ${\cal O}_{1}^{\mu \nu}$ from (\ref{a})
	\begin{equation}
	{\cal O}_{1}^{\mu \nu}=\frac{1}{c_1} \left[^{<}D \sigma^{\mu \nu \, >} + T \nabla_{\lambda}\left(\frac{u^{\lambda}}{4 T}\right) \sigma^{\mu \nu}- c_2 {\cal{O}}_{2}^{\mu \nu}-........- c_9 {\cal{O}}_{9}^{\mu \nu}\right]  \label{b}
	\end{equation}
	and substitute it i,e. (\ref{b}) in (\ref{3}) to obtain,
	\begin{eqnarray}
	\pi^{\mu \nu}&=& -2\eta \sigma^{\mu \nu} +\frac{\lambda_1}{c_1}\, ^{<}D \sigma^{\mu \nu \, >}+\frac{\lambda_1}{c_1} T \nabla_{\lambda}\left(\frac{u^{\lambda}}{4 T}\right) \sigma^{\mu \nu} + \Sigma_{i=2}^{i=9} \left(\lambda_{i}-\frac{c_i}{c_1}\lambda_1 \right){\cal O}^{\mu \nu}_{i}  \label{Navier-general}
	\end{eqnarray}
 To actually reproduce the relaxation-type equation up to second order in gradients, we finally substitute the first-order solution\footnote[1]{ As $D\sigma^{\mu \nu}$ and $D\pi^{\mu \nu}$ are both second order of gradients, it is possible to substitute the first order results of dissipative fluxes in second order theories. Therefore one can switch between $D\pi^{\mu \nu}$ and $D(-2\eta \sigma^{\mu \nu})$ thereby connecting to the MIS theory without loss of accuracy at the second order, see for eg. \cite{Romatschke-2}. } i,e. $\sigma^{\mu \nu}=-\frac{1}{2\eta} \pi^{\mu \nu}$ in all second-order terms of  eq.(\ref{Navier-general}) to  obtain,
	\begin{eqnarray}
	\tau_{\pi} \,^{<}D \pi^{\mu \nu \, >} + \pi^{\mu \nu}&=& -2\eta \sigma^{\mu \nu}+\tau_{\pi}\left[ \frac{D \eta}{\eta}  - \frac{(\nabla.u)}{4}   
	+ \frac{D\ln T}{4} \right] \pi^{\mu \nu} \nonumber \\
	& +& \xi_2 \,\mathcal{D}^{<\mu}_\bot \mathcal{D}^{\nu>}_\bot \ln T + \xi_3\, (\mathcal{D}^{<\mu}_\bot \mu) (\mathcal{D}^{\nu>}_\bot \mu) + \xi_4\,\mathcal{D}^{<\mu}_\bot \mathcal{D}^{\nu>}_\bot \mu +\xi_5 \,\mathcal{D}^{<\mu}_\bot \ln T \mathcal{D}^{\nu>}_\bot \mu  \nonumber \\
	&- &\ \frac{\xi_6}{2 \eta} \,\pi^{\gamma<\mu}\Omega^{\nu>}\,_{\gamma} + 
	\frac{\xi_7}{4\eta^2} \,\pi^{\gamma<\mu}\pi^{\nu>}\,_{\gamma} +
	\xi_8 \, \Omega^{\gamma<\mu}\Omega^{\nu>}\,_{\gamma}   \nonumber \\
	&+&\kappa_1 \,	R^{< \mu \nu>}+
	\kappa_2 \,u_{\alpha} u_{\beta} R^{\alpha<\mu \nu> \beta} 
	\label{Navier-Eckart} 
	\end{eqnarray}
where $\tau_{\pi}= \displaystyle \frac{\lambda_1}{2c_1 \eta}$ is the transport coefficient known as relaxation-time coefficient for $\pi^{\mu \nu}$ which preserves causality in the Navier-Stokes equation in absence of bulk viscosity, heat flux and charge diffusion. 
	Following a similar procedure and systematically considering the second order terms, the structure of bulk viscosity up to second order with relaxation time coefficient $\tau_{\Pi}$ becomes,
	\begin{eqnarray}
	\Pi&=&- \zeta (\nabla .u) - \tau_{\Pi}D(-\zeta\nabla.u) \nonumber \\
	&&+ \zeta_2\, (\mathcal{D}^{\alpha}_\bot \ln T)\mathcal{D}_{\alpha}^\bot \ln T +
	\zeta_3\,(\nabla.u)^2 + \zeta_4\, R
	+\zeta_5\, u^{\alpha}u^{\beta}R_{\alpha \beta} \nonumber \\
	&&+\zeta_6 \sigma^{\alpha \beta}\sigma_{\alpha \beta}+
	\zeta_7\,\Omega^{\alpha \beta} \Omega_{\alpha \beta}+
	\zeta_8\, (\mathcal{D}^{\alpha}_\bot \mu) \mathcal{D}_{\alpha}^\bot \mu +
	\zeta_9\,\mathcal{D}_{\alpha}^\bot\mathcal{D}^{\alpha}_\bot \mu +
	\zeta_{10}\,(\mathcal{D}_\bot^{\alpha}\mu)\mathcal{D}_{\alpha}^\bot \ln T \label{Final-bulk}
	\end{eqnarray}
	and in the same way heat-flow vector may be written as,
	\begin{eqnarray}
	q^{\mu}&=&-\kappa \left(\mathcal{D}^{\mu}_\bot T+T Du^{\mu}\right)+\tau_{q}D\left[\mathcal{D}^{\mu}_\bot T+T Du^{\mu}\right] \nonumber \\
	&& +\chi_{2}\,(\mathcal{D}_{\alpha}^\bot \ln T)\sigma^{\alpha \nu}+\chi_{3}\,(\mathcal{D}_{\alpha}^\bot \ln T)\Omega^{\alpha \nu}+\chi_{4}\,(\mathcal{D}_{\alpha}^\bot \mu)\displaystyle \frac{(\nabla.u)}{3} \triangle^{\alpha \nu }\nonumber \\
	&&+\chi_{5}\,(\mathcal{D}^{\nu}_\bot \mu) (\nabla.u)+
	\chi_6\,(\mathcal{D}_{\alpha}^\bot \mu)\sigma^{\alpha \nu}+
	\chi_7\,\triangle^{\alpha \nu} u^{\gamma} R_{\alpha \gamma} +
	\chi_8\,\mathcal{D}_{\alpha}^\bot \sigma^{\alpha \nu} \nonumber \\
	&&+ \chi_9\,\mathcal{D}_{\alpha}^\bot \Omega^{\alpha \nu}+
	\chi_{10}\,(\mathcal{D}_{\alpha}^\bot \mu)\Omega^{\alpha \nu} + \chi_{11}\, (\mathcal{D}_{\alpha}^\bot \mu)R^{\alpha \nu} \label{Final-heat}
	\end{eqnarray}
	where $\tau_{q}$ is the relaxation time coefficients of heat flux. Each of these measure the time the system takes to return to the equilibrium state in absence of dissipative quantities.  
	Thus eq.(\ref{Final-shear}), eq.(\ref{Final-bulk}) and eq.(\ref{Final-heat}) are respectively the general forms of shear viscosity tensor, bulk viscosity and heat flow vector in the Eckart frame up to second order in gradient expansion. 
	Moreover, one can also show that relaxation-type dynamical equations can be constructed for heat flux vector $q^{\mu}$ and bulk viscosity $\Pi$ as well.
	There are in total nine second order transport coefficients namely $\tau_{\pi}, \xi_{2}, \xi_{3}, \xi_{4},\xi_{5}, \xi_{6},\xi_{7}$, $\kappa_1$ and $\kappa_2$ for the shear viscosity. 
	Together with $\tau_{\Pi}$, there are ten second order transport coefficients i,e. $\zeta_1, \zeta_2, \zeta_3,\zeta_4,\zeta_5,\zeta_6,\zeta_7,\zeta_8, \zeta_9$ and $\zeta_{10}$ corresponding to the bulk viscosity. 
	The transport coefficients corresponding to the heat flow count to total eleven in number which are given by $\tau_{q},\xi_2,\xi_3,\xi_4,\xi_5,\xi_6,\xi_7,\xi_8,\xi_9,\xi_{10},\xi_{11}$.
	\section{Discussions on causality aspects: determination of characteristic velocities}
	In this section we will study the causality of second order hydrodynamics of non-ideal fluids in Eckart frame using gradient expansion scheme. It is well-known that the relativistic Navier-Stokes equation formulated in the context of first order theories give rise to superluminal propagations in the small wavelength limit, are unstable \cite{Hiscock,Hiscock-2} and lead to obstacles in posing initial value problems. Hence one seeks to incorporate second order gradients and introduce relaxation time coefficients for each of the dissipation variables.
	Before proceeding further, we note that in the direction of investigating the causal behavior, a complete set of hyperbolic system of first order PDEs for non-ideal fluids has been formulated in the $3+1$ representation of Einstein equations \cite{Peitz-1} and the hyperbolic property of the MIS theory is extensively studied \cite{Peitz-2}. \\
	The gradient expansion scheme is based on the idea that hydrodynamics of non-ideal fluids can be constructed in a systematic way by considering gradients of hydrodynamical variables and the source $g_{\mu \nu}$ order by order under the condition that the system is slightly out of equilibrium. Applied to the hydrodynamics which is an effective theory, the gradients and the fluctuations are small. 
	In this sense, the Navier-Stokes equation is first order theory involving single gradients in the constitutive relations. The second order theory is formulated by systematically considering all possible second order gradient terms for constructing the general structures of dissipatiive flux quantities, at the same time also embodies MIS theory with additional transport coefficients. Nevertheless, second order gradients are small compared to first order gradients and it is the out-of -equilibrium fluid dynamics, when the gradients can longer be considered to be small. This situation is beyond the scope of the current work. 
	In the present work, we restrict ourselves to the regime of fluid dynamics when the gradients are small and is legitimate to describe near-equilibrium systems.\\
	 First formulated in Landau frame with no conserved charges, the method is proved to give rise to a  theory with finite propagation speeds in presence of linear fluctuations of fluid hydrodynamical variables and the metric \cite{Romatschke-1}. In order to establish that present formalism of second order hydrodynamics leads to a causal theory, we will determine velocity of propagation of the fluctuations. For this, we determine the collective modes of Navier-Stokes equation in presence of second order gradients by employing the variational approach in which sources directly couple to the energy momentum tensor. As the metric is the source for the energy momentum tensor, the retarded two-point correlator in the Minkowski spacetime is given by,
	\begin{equation}
	G^{\mu \nu, \rho \sigma} = -2\frac{\delta T^{\mu \nu}(t,\textbf{x},g)}{\delta g_{\rho \sigma}}\Big{|}\,_ {\delta g_{\mu \nu}=0}  \label{corr}
	\end{equation}
	which is evaluated by expanding the metric up to linear order in perturbation around Minkowski spacetime as $g_{\mu \nu}= \eta_{\mu \nu}+\delta g_{\mu \nu}$. 
	The  metric fluctuations $\delta g_{\mu \nu}$ act as source to induce perturbations in energy density $\delta e$, pressure $\delta p$, temperature $\delta T$ and particle number $ \delta n$, all of which are proportional to $\delta g_{\mu \nu}$ around a constant background.\\
As dissipation arises as a result of small departures from the static equilibrium configuration, it is sufficient for us to investigate the causality of the Navier-Stokes equation by studying the collective modes of the relativistic hydrodynamics up to linear order in perturbations.
 All effects steaming from dissipation are then captured by analyzing linear perturbations around equilibrium densities and temperature. Up to linear order in fluctuations, the energy momentum tensor and particle current are given by,
\begin{eqnarray}
\delta T^{\mu \nu}& =&  u^{\mu }_{(0)}u^{\nu}_{(0)}\delta e+p_0\delta g^{\mu \nu}+(\delta P+\delta\Pi)\triangle^{\mu \nu}_{(0)}+\delta(q^{\mu}u^{\nu}+q^{\nu}u^{\mu}) +\delta\pi^{\mu\nu} \\
\delta N^{\mu}&=& u^{\mu}_{(0)}\delta n+ n_0\delta u^{\mu}  
\end{eqnarray}
where $e_{0}$, $p_{0}$ and $n_{0}$ are equilibrium values of energy density, pressure and number density respectively and $u_{(0)}= (1,0)$. In order to determine the retarded correlation function each of $\delta \Pi$, $\delta \pi^{\mu \nu}$ and $\delta q^{\mu}$ are expanded in $\delta g_{\mu \nu}$ and without loss of generality, we choose specific co-ordinate dependence of the metric namely, $\delta g_{\mu \nu}\equiv \delta g_{\mu \nu}(t,z)$ for calculational simplifications. Considering the equation of state as $p \equiv p(e,n)$ and $T \equiv T(e,n)$, up to linear order in perturbations, the hydrodynamical variables are expanded as,
\begin{eqnarray}
e&=&e_{0}+ \delta e, \qquad n= n_{0}+ \delta n, \qquad
u^{\mu}=u^{\mu}_{(0)}+\delta u^{\mu} \\
p&=&p_{0}+ \alpha_{pn}\delta n + \alpha_{pe}\delta e,  \label{exp-p}\qquad \,
T=T_{0}+ \alpha_{Tn}\delta n + \alpha_{Te}\delta e
\end{eqnarray}
where the coefficients in eq.(\ref{exp-p}) are given by,
\begin{eqnarray}
\alpha_{pn}= \frac{\partial p}{\partial n},\quad\alpha_{pe}= \frac{\partial p}{\partial e}, \quad \alpha_{Tn}= \frac{\partial T}{\partial n}, \quad\alpha_{Te}= \frac{\partial T}{\partial e}
\end{eqnarray}
For small perturbations, one can safely ignore non-linear terms in heat flux, shear viscosity tensor and bulk viscosity. Then, we have,
\begin{eqnarray}
	\delta \pi^{\mu \nu}&=&-2\eta\delta \sigma^{\mu \nu}- \tau_{\pi} \delta D\pi^{<\mu \nu>} + \kappa_1 \,\delta R^{< \mu \nu>}+
	\kappa_2 \delta (\,u_{\alpha} u_{\beta} R^{\alpha<\mu \nu> \beta})\\ \nonumber
	\delta \Pi^{\mu \nu}&=&-\zeta \delta (\nabla.u)-\tau_{\Pi} \delta (D\Pi) \\ \nonumber
	\delta q^{\mu}&=&- \kappa  \delta(\mathcal{D}^{\mu}_\bot T+T Du^{\mu}t)-\tau_{q} \delta (D q^{\mu})
\end{eqnarray}
Using Fourier ansatz for perturbations,
\begin{eqnarray}
\delta e&=&e^{-i\omega t+ikz}k_{\delta e},\quad \delta n=e^{-i\omega t+ikz}k_{\delta n}, \nonumber\\  
\delta T&=&e^{-i\omega t+ikz}(\alpha_{Tn}k_{\delta n} + \alpha_{Te}k_{\delta e}), \quad \delta p= e^{-i\omega t+ikz}(\alpha_{pn}k_{\delta n} + \alpha_{pe}k_{\delta e}), \nonumber\\
\delta g_{\mu \nu}(t,z)&=&e^{-i\omega t+ikz}k_{h_{\mu \nu}}
\end{eqnarray}
one can compute the two-point correlators to determine the collective modes using eq.(\ref{corr}) in the momentum space using the variational approach in which the quantities $G^{\mu \nu, \rho \sigma}$ are calculated by by taking variation of one-point functions with respect to sources. For example, 
\begin{equation}
G^{xy,xy} = p_0 - \frac{i\eta \omega}{(1-i \omega \tau_{\pi} )}+ \frac{\omega^2 \kappa_1+\kappa_2(\omega^2+k^2)}{2(1-i \omega \tau_{\pi} )} 
\end{equation}
where $p_0$ is the contact term. The characteristic velocity of propagation of the shear mode can be determined from $G^{tx,tx}(\omega,k)$. In the hydrodynamic limit (i,e, the small $k$ limit), the shear mode channel is given by,
\begin{equation}
G^{tx,tx}(\omega,k)=e_0 + \frac{k^2(e_0+p_0) \left[2 \eta+i \omega(\kappa_1+\kappa_2)\right]}{-2 \eta k^2-2i \omega(1-i\omega\tau_{\pi})}
\end{equation}
where $e_0$ is the contact term. The singularities of $G^{tx,tx}$ will determine the shear modes and the dispersion relation is given by,
\begin{equation}
\eta k^2+i \omega(1-i \omega \tau_{\pi})=0
\end{equation}
which can be solved to obtain,
\begin{equation}
\omega^{\pm}=\displaystyle\pm\frac{-i\pm\sqrt{\displaystyle\frac{4 k^2 \tau_{\pi}\eta}{(e_0+p_0)}-1}}{2 \tau_{\pi}}
\end{equation}
and the group velocity is given by,
\begin{equation}
v_{g}^{shear}=\Big{|}\frac{d \omega^{+}}{dk}\Big{|}= \frac{2k \eta/h}{\sqrt{\displaystyle\frac{4 k^2 \tau_{\pi}\eta}{(e_0+p_0)}-1}} \label{shear-speed}
\end{equation}
The maximum velocity of propagation is then given by,
\begin{equation}
v_{max}^{shear}= lim_{k \rightarrow \infty}v_{g}^{shear}= \sqrt{\frac{\eta/(e_0+p_0)}{ \tau_{\pi}}}
\end{equation}
which is independent of $k$ and remains finite provided the relaxation time coefficient of the shear viscosity fulfills the condition $\tau_{\pi}> \eta/(e_0+p_0)$. The finite velocity of propagation of shear modes therefore preserves the causality of the relativistic Navier-Stokes equation in the Eckart frame.
On the other hand, the sound mode channel is given by,
\begin{eqnarray}
G^{tt,tt}&=&-2e_0-\frac{A(\omega,k)}{B(\omega,k)}\\[1mm]
\text{where,}A(\omega,k)&=&\left[3k^2(e_0+p_0)(i+\omega \tau_{\pi})(i+\omega \tau_{\Pi})-2k^4(\kappa_1+\kappa_2)(1-i\omega \tau_{q})+3ik^4(i+\omega \tau_{\pi})(2\xi_5-\xi_6)\right] \nonumber \\
&&\times \left[k^2 n_0 \kappa \alpha_{Tn}+\omega(e_0+p_0)(i+\omega \tau_{q})\right]
\end{eqnarray}
The dispersion relation is given by,
\begin{eqnarray}
B(\omega,k)=i\omega-i k^2 \frac{c_s^2}{\omega}-k^2 \left[\frac{4}{3(e_0+p_0)}\frac{\eta}{1-i \tau_{\pi}\omega}+\frac{\zeta}{(e_0+p_0)}\frac{1}{1-i \tau_{\Pi}\omega}\right]\\[2mm] \nonumber
+ k^2\frac{\kappa n_0}{(e_0+p_0)}\frac{(\alpha_{pn}\alpha_{Te}-\alpha_{pe}\alpha_{Tn})}{\omega^2(1-i \omega \tau_{q})}+k^2\frac{\kappa n_0}{(e_0+p_0)} \frac{\alpha_{Tn}}{(1-i \omega \tau_{q})}\\[2mm] \nonumber
+ k^4 \frac{\kappa\, \alpha_{Te}}{\omega}\left[\frac{4}{3(e_0+p_0)}\frac{\eta}{1-i\tau_{\pi}\omega}+\frac{\zeta}{(e_0+p_0)}\frac{1}{1-i\tau_{\Pi}\omega}\right]=0
\end{eqnarray}
where the speed of the sound is $c_s=\sqrt{\alpha_{pe}+\displaystyle\frac{n_0}{e_0+p_0}\alpha_{pn}}$. In the large $\omega$ limit, the solution of the above equation leads to,
\begin{eqnarray}
\omega^{\pm}=\pm k \sqrt{c_s^2+\frac{4\eta}{3\tau_{\pi}(e_0+p_0)}+\frac{\zeta}{\tau_{\Pi}(e_0+p_0)}-\frac{n_0 \kappa }{\tau_{q}(e_0+p_0)}\alpha_{Tn}}
\end{eqnarray}
and the maximal speed of propagation of sound mode is found to be,
\begin{equation}
v_{max}^{sound}=\sqrt{c_s^2+\frac{4\eta}{3\tau_{\pi}(e_0+p_0)}+\frac{\zeta}{\tau_{\Pi}(e_0+p_0)}-\frac{n_0 \kappa }{\tau_{q}(e_0+p_0)}\alpha_{Tn}}  \label{sound-speed}
\end{equation} 
which is independent of wavenumber $k$ and finite provided we have $\tau_{\pi}>\frac{4\eta}{(e_0+p_0)}$, $\tau_{\Pi}>\frac{\zeta}{(e_0+p_0)}$ and $\tau_{q}>\frac{n_0 \kappa \alpha_{Tn}}{(e_0+p_0)}$. All the transport coefficients at the first and the second order can be determined from Kubo formula in the context of an underlying microscopic theory. The finiteness of propagation velocities of the shear mode and the sound mode thus establish that second order theory of dissipative relativistic hydrodynamics constructed using the gradient expansion scheme in the Eckart frame preserve causality and fluctuations giving rise to  dissipative effects travel at a finite speed\footnote{In the same sense as MIS theory}. A crucial consideration of the presented formalism (which is at par with MIS theory.) is that dissipation arises as a result of small departures from the local equilibrium configuration of the system. These fluctuations are smaller compared to their respective equilibrium values. Hence we have considered linearized theory  of hydrodynamics where all fluctuations of hydrodynamical variables giving rise to viscous hydrodynamics are of linear order. We further note here eq.(\ref{sound-speed}) exactly reduces to the characteristic velocity of sound mode in absence of conserved charges found in \cite{Romatschke-3}.

	\section{Outlook and future possibilities}	
		
	\begin{itemize}	
	\item \textit{Accretion disks around black holes:}
	Recently, Event horizon telescope Collaboration (EHT) \cite{Akiyama} has published the first image of a rotating black hole located at the centre of M87 galaxy. In this image, the black hole can be seen to be surrounded by a disk-like luminous structure known as the accretion disk. 
	A key mechanism involved within the disk is the inward mass transport of the disk-matter which is accompanied by an outward angular momentum transport. Known as an accretion process/flow, one of the governing mechanisms of this process is thought to be the viscosity. 
	Moreover, other dissipative quantities like heat flux, vorticity might also come into play. 
	Under this situation, a novel aspect is, our formalism developed in the Eckart frame lends its support for understanding the origin of viscosity within an accretion disk in presence of curvature effects through Ricci tensor or ricci scalar of a given black hole spacetime (see, eq.(\ref{Final-shear})) in addition to having causal Navier-Stokes equation and heat conduction equation . 
	When applied either for stationary configurations or dynamical cases, our approach  illustrates imprints of curvature effects in the strong gravity regime on different properties of accretion disks through general forms of bulk viscosity, shear viscosity and the heat flow. 
	In addition, the non-linear terms of the shear tensor, vorticity may also play their roles which so far has not been considered in the literature.
	As a first application of our formalism, we have chosen the thick accretion disk. In \cite{Lahiri}, we have incorporated effects of causality preserving term and curvature effects to study the shape of the accretion disk in Schwarzchild spacetime which marks the first-ever step towards comprehending the role of viscosity and curvature in Polish doughnuts.
	Further investigation involves studying impacts of curvature effects through viscosity on the properties of thin accretion disk in a given black hole spacetime. This work will be reported shortly. 
	So far the effects of viscosity in the realm of $\alpha$ viscosity formalism has been considered for thin accretion disks, ADAFs \cite{Gammie}.  
	Our formalism can effectively be applied in different families of accretion disks subjected other forms of interactions like magnetic field, charge in a given black hole background.
	\item \textit{Static equilibrium  configuration : Tolmann}--Oppenheimer-Volkoff (TOV) stars: The set of TOV equation describes the spherically symmetric static equilibrium configuration of a star supported by an ideal fluid for a given equation of state \cite{Tolman}, in particular with the equation of state of degenerate Fermi gas of neutrons were studied in \cite{Oppen}.
	The TOV equations are valid for stars in hydrostatic equilibrium where the inward gravitational pull is balanced by the pressure of the fluid.
	In the static spherically symmetric configuration, the most general form of metric contains two unknown mass function $m(r)$ and the potential function $\Phi(r)$ and the four velocity of the fluid in the static situation contains only the time component. 
	Then with the given equation of state that relates the fluid pressure $p(r)$ with its energy density $e(r)$ where $r$ is the radial co-ordinate, the internal structure of the star is determined by solving the Einstein equations for the unknowns functions $m(r)$, $\Phi(r)$ and $p(r)$.\\
	In the present context, it will be interesting to study the hydrostatic equilibrium condition in presence of  bulk viscosity which is responsible for the volume expansion. 
	As an initial step, it is worthwhile to consider the Ricci scalar terms from eq.(\ref{Final-bulk}). In this context the effective fluid pressure may be given by $P(r)=p(r)+ \Pi(r)$ where $\Pi(r)=\zeta_4\, R
	+\zeta_5\, u^{\alpha}u^{\beta}R_{\alpha \beta}$. 
	We note here that due to staticity and spherical symmetry, the first order quantity of the bulk viscosity vanishes i,e. $\nabla .u=0$.
	It will be then worthwhile to investigate the interplay of the curvature terms appearing in Einstein tensor and in the energy momentum tensor due to bulk viscosity that will will help the star to reorganize itself in presence of the effective pressure $P$. 
	In order to compute changes in the radius of the star due to curvature terms of bulk viscosity, the modified TOV equations are required to be solved numerically with initial value as that of ideal fluid to estimate the radius of the star where the pressure vanishes. 
	The modified radius of the star, in which the star settles to, will give rise to its new hydrostatic equilibrium configuration with small corrections arising due to curvature effects of bulk viscosity.
	For example, in a realistic neutron star, these effects are present but are quite small, in fact the modifications in the radii are suppressed of the order of $GeV^2/M_{pl} << 1$. 
	On the other hand, it might be useful to expand the radius of the star in Taylor series around its equilibrium value and consider terms linear in the radius.
	\item \textit{Physics of early universe}: In \cite{Lahiri-2}, it is shown that curvature effects in bulk viscosity arising in the context of second order gradient method solely gives rise to the exponential expansion of the early universe.
	Similar considerations can be considered in the Eckart frame description for determining cosmological solutions supported by Ricci terms, however, the solutions will be characterized by transport coefficients of the Eckart frame.
	The fluid four-velocity in the Eckart frame is taken along the flow of conserved charge in the system, as a result, the charge flow vanishes. Consequently, the continuity equation remains unchanged in absence of charge diffusion.
		This brings us to the condition that, in the Eckart frame, we are unable to investigate impacts of non-zero charge diffusion on physics of early universe.\\
		However, the effects of non-constant chemical potential on the physics of the early universe  can be captured due to the term $\zeta_9\,\mathcal{D}_{\alpha}^\bot\mathcal{D}^{\alpha}_\bot \mu$ in the bulk viscosity. In that case, the equation of state may be taken as $p \equiv p(T,n)$ and the effective fluid pressure becomes $p+\Pi$ where $\Pi$ given by eq.(22).
		In the context of \cite{Tawfik:2011sh}, using the bulk viscosity transport coefficient $\zeta$ evaluated using lattice QCD calculations and relaxation time coefficient $\tau_{\Pi}$, one can determine the effects of chemical potential on the cosmological solutions (including the de-Sitter universe) by solving the continuity equation, modified Friedmann equation and the evolution equation of the bulk viscosity which is given by,
		\begin{eqnarray}
		\tau_{\Pi}D\Pi + \Pi&=&- \zeta (\nabla .u) 
		+ \zeta_2\, (\mathcal{D}^{\alpha}_\bot \ln T)\mathcal{D}_{\alpha}^\bot \ln T +
		\zeta_3\,(\nabla.u)^2 + \zeta_4\, R
		+\zeta_5\, u^{\alpha}u^{\beta}R_{\alpha \beta} \nonumber \\
		&+&\zeta_6 \sigma^{\alpha \beta}\sigma_{\alpha \beta}+
		\zeta_7\,\Omega^{\alpha \beta} \Omega_{\alpha \beta}+
		\zeta_8\, (\mathcal{D}^{\alpha}_\bot \mu) \mathcal{D}_{\alpha}^\bot \mu +
		\zeta_9\,\mathcal{D}_{\alpha}^\bot\mathcal{D}^{\alpha}_\bot \mu +
		\zeta_{10}\,(\mathcal{D}_\bot^{\alpha}\mu)\mathcal{D}_{\alpha}^\bot \ln T  \nonumber
		\end{eqnarray}
		 It will be interesting to investigate how chemical potential and of curvature terms bring about changes on the critical temperature $T_{c}$ and on first-order quark-hadron phase transitions. 
		Much in the similar lines of \cite{Tawfik:2010bm}, it will be worth analyzing the implications of the second order theory using gradient expansion scheme in presence of non-zero chemical potential on the behavior of the scale factor during the quark phase and hadron phase. This will involve studies the evolution of the temperature of the universe, nature of Hubble expansion, energy density.
\end{itemize}
	The present work is an initial step towards presenting a causal theory of hydrodynamics of relativistic non-ideal fluids using gradient expansion scheme in the Eckart frame up to second order in compliance with the theory of general relativity. The immediate next step would be to compute the transport coefficients explicitly in this frame. For instance, it has been suggested in \cite{Monnai} that the second order transport coefficients in this frame are related to that of the Landau frame under certain circumstances. This would be a nice to study the frame dependent nature of transport coefficients if any. It would also be worth investigating how many of these second order transport coefficients are really independent, following the formalism developed in \cite{Romatschke-2, Bhatta}. 
	We leave these issues for future work.

	\section*{ Acknowledgement}
	 I sincerely thank Paul Romatschke for illuminating suggestions and discussions at various stages of this work. I am also thankful to Claus Laemmerzahl for some discussions. I am thankful to anonymous referees for providing valuable suggestions and remarks that contributed to the improvement of the manuscript.
	 This work is supported by Deutsche Forschungsgemeinschaft (DFG) with grant/40401089.

\end{document}